\documentclass[12pt,preprint]{aastex}

\shorttitle{ IRAS 17423$-$1755 (Hen 3$-$1475) revisited}
\shortauthors{Manteiga et al.}

\begin{document}

\title{ IRAS 17423$-$1755 (HEN 3$-$1475) REVISITED:\\
    AN O$-$RICH HIGH$-$MASS POST$-$ASYMPTOTIC GIANT BRANCH STAR}

\author{M. Manteiga}
\affil{ Departamento de Ciencias de la Navegaci\'on y de la Tierra. 
Universidade da Coru\~na. Paseo de Ronda 51, E-15011 A Coru\~na, Spain }
\email{manteiga@udc.es}

\author{ D. A. Garc\'\i a-Hern\'andez\altaffilmark{1}}
\affil{Instituto de Astrof\'\i sica de Canarias (IAC), V\'\i a Lactea s/n, 
E-38200 La Laguna, Tenerife, Spain}

\author{A. Ulla}
\affil{ Departamento de F\'\i sica Aplicada. Campus Lagoas-Marcosende. 
Universidade de Vigo. E-36310 Vigo, Pontevedra, Spain }

\author{ A. Manchado\altaffilmark{1,2}}
\affil{Instituto de Astrof\'\i sica de Canarias (IAC), V\'\i a Lactea s/n, 
E-38200 La Laguna, Tenerife, Spain}

\author{P. Garc\'\i a- Lario}
\affil{ Herschel Science Centre. European Space Astronomy Centre (ESAC)/ 
European Space Agency (ESA). Villafranca del Castillo. Apartado de Correos 78. E-28080 Madrid, Spain}

\altaffiltext{1}{ Departamento de Astrof\'\i sica, Universidad de La Laguna (ULL), 
E-38205 La Laguna, Tenerife, Spain }
\altaffiltext{2}{ Consejo Superior de Investigaciones Cient\'\i ficas, Spain }

\begin{abstract}
The high-resolution ({{\it{R}}$\sim$600) {\it{Spitzer}}/IRS spectrum of the bipolar proto
planetary nebula (PN) IRAS 17423$-$1755 is presented in order to clarify the
dominant chemistry (C-rich versus O-rich) of its circumstellar envelope as well as
to constrain its evolutionary stage. The high quality {\it{Spitzer}}/IRS spectrum shows
weak 9.7 $\mu$m absorption from amorphous silicates. This confirms for the first time the O-rich nature of
IRAS  17423$-$1755 in contradiction to a previous C-rich classification, 
which was based on the wrong identification of the strong
3.1 $\mu$m absorption feature seen in the {\it{Infrared Space Observatory}} (\it{ISO}})
spectrum as due to acetylene (C$_{2}$H$_{2}$). The high-resolution {\it{Spitzer}}/IRS
spectrum displays a complete lack of C-rich mid-IR features such as molecular
absorption features (e.g., 13.7 $\mu$m C$_{2}$H$_{2}$, 14.0 $\mu$m HCN, etc.) or
the classical polycyclic aromatic hydrocarbon infrared emission bands.
Thus, the strong 3.1 $\mu$m absorption band toward IRAS 17423$-$1755 has to be
identified as water ice. In addition, an [NeII] nebular emission line at 12.8
$\mu$m is clearly detected, indicating that the ionization of its central region
may be already started. The spectral energy distribution in the infrared
($\sim$2$-$200 $\mu$m) and other observational properties of IRAS
17423$-$1755 are discussed in comparison with  the similar post-asymptotic giant branch (AGB) objects IRAS
19343$+$2926 and IRAS 17393$-$2727. We conclude that IRAS 17423$-$1755 is an
O-rich high-mass post-AGB object that represents a link between OH/IR
stars with extreme outflows and highly bipolar PN. 

\end{abstract}


\section{Introduction}
IRAS 17423$-$1755 (Hen 3$-$1475) was first suggested by \citet{ParthaPottasch89} as
a possible member of the transition phase from the asymptotic giant branch (AGB)
to the planetary nebula (PN) stage due to its unusual IRAS colors. The high
values of the [NII]/H$\alpha$ ratios in the outflowing material detected by
\citet{Riera95} and the low luminosity deduced for the central star allowed them
to confirm the classification of this object as an evolved star.  {\it{Hubble Space
Telescope}} ({\it{HST}}) and {\it{Very Large Array}} ({\it{VLA}}) observations by \citet{Bobrowski95}
showed the presence of both OH maser emission and highly collimated ionized 
outflows like those detected in OH/IR stars or very young PN. 

 {\it{HST}} images revealed a rich and complex morphological structure in the
circumstellar material (see Section 4 for details). The outflow is collimated in bipolar jets along several
condensations of shock-excited gas that extend about 11 arcsec. The lobes
show expansion velocities of about 425 km s$^{-1}$ and a high velocity jet
($\sim$900 km s$^{-1}$) in the inner part of the lobes \citep{Riera95}. The nebula displays a
remarkable point symmetry that has been interpreted as due to the precession of
a central binary system that undergoes episodic events of mass loss.
\citet{Bobrowski95} proposed that the expanding shell has a torus-like structure
where the OH emission originates in a high density region, where H$_{2}$O is
dissociated and further collimated in the observed jets. Additionally, 
\citet{SanchezContreras01} found evidences of ultrafast winds (up to 2300 km s$^{-1}$) highly collimated 
and located close to the central star which could be a relatively young post-AGB outflow not strongly altered by interaction with the AGB.

Stars at the end of the AGB phase are characterized by severe mass loss
($10^{-8}$ to $10^{-4}$ M$_{\odot}$ yr$^{-1}$), which results in the formation of circumstellar
envelopes \citep{Herwig05}. The spherical symmetry of the envelopes of AGB stars
is translated into a variety of shapes in the PN phase by a mechanism or mechanisms
not as yet well understood. There is increasing evidence that at least in some 
instances the shaping starts at the end of the AGB phase \citep{vanWinckel03}. 
IRAS 17423$-$1755 is a spectacular example that may
represent a link between OH/IR stars with extreme outflows and highly bipolar
PN.

The spectral energy distribution (SED) of extreme (e.g., highly embedded) OH/IR AGB
stars is characterized by the presence of strong and broad amorphous silicate
absorption features at 9.7 and 18 $\mu$m together with crystalline silicate
absorption/emission features from 10 to 45 $\mu$m \citep{Sylvester99,GarciaH07}.  At
the end of the AGB phase, the crystalline silicate features become dominant and can be
observed in more evolved O-rich PN \citep{Molster01}. Comparison of the
 {\it{Infrared Space Observatory}} ({\it{ISO}}) observations of O-rich dust shells surrounding
evolved stars with laboratory data suggested the presence of several families of
crystalline silicates, such as olivines and pyroxenes, and marked the beginning of an
emerging discipline: the mineralogy of stellar and other astronomical (i.e., cometary)
dust shells. Water ice features at 3.1, 43, and 62 $\mu$m have been additionally
observed in heavily obscured and extremely bipolar sources such as the post-AGB star IRAS
19343$+$2926 (or M1$-$92, see e.g. Dijkstra et al. 2006 and references therein).

\citet{Gauba04} studied the {\it{ISO}} spectra of seven hot post-AGB stars including IRAS
17423$-$1755. DUSTY models \citep{Ivezic97} were fitted to optical, near- and far-infrared (IRAS
and {\it{ISO}}) photometry in order to reconstruct the SEDs and to derive physical
parameters such as dust temperatures, mass loss rates, angular radii and the inner
boundary of the dust envelopes. For the particular case of IRAS 17423$-$1755
they considered a combination of silicates and carbon in the circumstellar
environment. They reported the presence of a broad absorption feature at 3.1
$\mu$m that they identified as due to the presence of C$_{2}$H$_{2}$ and/or HCN
in the circumstellar envelope. This identification led these authors to infer a
C-rich chemistry for the shell. More recently, \citet{Cerrigone09} presented
observations of a sample of 26 hot post-AGB stars with the Infrared Array
Camera and the Infrared Spectrograph (IRS) on board the  {\it{Spitzer Space
Telescope}}. These observations were analyzed together with Two Micron All Sky
Survey, IRAS and radio centimeter data in order to model the SEDs of the
targets. \citet{Cerrigone09} classified IRAS 17423$-$1755 as a C-rich star on
the basis of the \citet{Gauba04} report of the C$_{2}$H$_{2}$ feature at 3.1
$\mu$m and in the absence of a strong 9.7 $\mu$m amorphous silicate
absorption/emission feature in their low-resolution (R$\sim$64$-$128)  {\it{Spitzer}}
spectrum. However, they pointed out that the expected polycyclic aromatic
hydrocarbon (PAH) features in the 5--12 $\mu$m region are not detected. It is to
be noted here that weak and narrow molecular absorptions from C-based molecules such
as 13.7 $\mu$m C$_{2}$H$_{2}$, 14.0 $\mu$m HCN, etc., are difficult to detect 
at the low resolution of their  {\it{Spitzer}} spectrum. The detection of these C-rich
molecular absorptions - typical of C-rich AGB/post-AGB stars - requires in most 
of the cases higher resolution observations such as those provided by the high-resolution 
modes of {\it{Spitzer}} ({\it{R}}$\sim$600) and {\it{ISO}} ({\it{R}}$\sim$1000) (see e.g., \citet{Cernicharo99,Cernicharo01,
GarciaH09}).

The controversial origin of the mid- to far-IR features in IRAS 17423$-$1755
merits a re-analysis of the dust features observed in the high-resolution
and higher quality {\it{Spitzer}} spectrum. In Section 2 we present the new {\it{Spitzer}}
observations together with the construction of the overall SED of the nebula as
observed by both {\it{Spitzer}} and {\it{ISO}}, while in Section 3 the evidence for
an O-rich chemistry is analyzed and discussed. The evolutionary stage of IRAS
17423$-$1755 is discussed, including our new results, in Section 4 while 
a summary of our main conclusions is presented in Section 5.


\section{ {\bfseries\itshape{Spitzer}} and  {\bfseries\itshape{ISO}} observations of IRAS 17423$-$1755 }

High-resolution (R$\sim$600)  {\it{Spitzer}}/IRS spectra of IRAS 17423$-$1755 are now
available in the   {\it{Spitzer}} public database. Short-high (SH: 9.9-19.6 $\mu$m) and
Long-high (LH: 18.7-37.2 $\mu$m) observations were obtained on 2009 April 21
under the General Observer Program $\#$50777 (P.I.: B. McCollum). A typical
signal-to-noise ratio (S/N) higher than 50 was easily reached by using just
three cycles of 6 s for both SH and LH modules. However, the S/N is much lower
for wavelengths longer than 34 $\mu$m - the red end of the LH module that is
affected by a strong noise level. The post-bcd products (one spectrum for each
nod position) automatically reduced by the IRS Custom Extractor (SPICE) were
retrieved from the  {\it{Spitzer}} database. The automatic data reduction includes the
extraction from the 2-dimensional images as well as the wavelength and flux calibration.
The  {\it{Spitzer}}-contributed software SMART \citep{Higdon04} was used for cleaning
residual bad pixels, spurious jumps and glitches and for smoothing and merging.
The short-low (SL: 5.2-14.5 $\mu$m)  {\it{Spitzer}} spectrum reported by
\citet{Cerrigone09}  was also retrieved from the  {\it{Spitzer}} database. We found a
good match ($\leq$5\%) between the SL and SH module spectra. However, the
absolute flux level of the LH module spectra was found to be $\sim$8 $\%$ higher
than the SH module spectra. Thus, we scaled the LH observations to the SH ones
in order to obtain the final high-resolution  {\it{Spitzer}} spectrum of IRAS
17423$-$1755. The good match between the  {\it{Spitzer}}/IRS SL and the SH spectra is 
illustrated in Figure 1.
For comparison, we also retrieved from the  {\it{Spitzer}}
database the high-resolution  {\it{Spitzer}} observations of the similar post-AGB stars
IRAS 19343$+$2926 and IRAS 17393$-$2727 (an OH/IR massive post-AGB already
reported by \citet{GarciaH07}). 

The full range {\it{ISO}} SWS+LWS spectrum of IRAS 17423$-$1755 was obtained in 1997 March as
part of the open-time program PGARCIA.PN on spectroscopy of proto-PN candidates. Both
SWS and LWS spectra were taken in the full scan AOT1 mode \citep{deGraauw96} at speed
1. The data were processed using the standard Interactive Analysis Software, version
7.0 of the SWS and LWS off-line processing system at the Max Planck Institute
for Extraterrestrial Astronomy (Garching, Germany); see \citet{GarciaLario99} for more
details on the reduction procedure. All detector signals were inspected for spurious
features, which were removed. No fringes were seen in the {\it{ISO}} spectrum of this
source. There is  good agreement between our SWS spectrum and IRAS low resolution
overlap region, and a good match between the SWS and LWS spectra. The 3.1 $\mu$m
absorption band can  be clearly observed in SWS AOT Band 1D. No firm conclusions can be
drawn concerning the presence of any other solid-state dust feature because of the
strong noise level in the {\it{ISO}} spectrum. The contribution from Galactic cirrus 
emission was estimated by IRSKY at the Galactic coordinates of the nebula.  It was
fitted by polynomials and subtracted from the resulting {\it{ISO}} spectra. The final
spectrum was transformed to standard FITS format and the subsequent processing and
analysis was performed using IRAF  routines.

Another post-AGB nebula sharing some observational properties with IRAS 17423$-$1755
is IRAS 19343$+$2926 (M1$-$92), the Minkowski Footprint \citep{Eiroa83}. Both nebulae
display similar morphological properties (bipolarity), a comparable effective
temperature for the central (B-type) star, host an OH maser and also present analogous
infrared spectral distributions (see also Section 4). In order to compare their infrared
properties, we have extracted the {\it{ISO}} spectra for this source from the archive and performed a
similar SWS+LWS reduction procedure as in the case of IRAS 17423$-$1755. The {\it{ISO}}
spectrum of IRAS 19343$+$2926 is of much higher quality than that of IRAS
17423$-$1755. Indeed, \citet{Kraemer02} have previously classified the {\it{ISO}} spectra of
IRAS 17423$-$1755 and IRAS 19343$+$2926 with the classes 5.F: and 5.SA, respectively.
This means that both spectra have red continua with the {\it{ISO}} spectrum of IRAS 17423$-$1755 
displaying no strong features
superimposed on the SED, and the {\it{ISO}} spectrum of IRAS 19343$+$2926 displaying a clear 10 $\mu$m 
silicate absorption feature and classified as an O-rich source. Objects in class 5.SA often 
also possess water ice, CO and CO$_{2}$ features.
Additionally, the IRS spectrum of the post-AGB star IRAS 17393$-$2727
\citep{GarciaH07}, will also be considered for comparison with IRAS 17423$-$1755.

We have constructed the SEDs of IRAS 17423$-$1755 and IRAS 19343$+$2926 in the
spectral region from 2 to $\sim$180 $\mu$m by using the IRS  {\it{Spitzer}} data in the
interval from 5 to 34 $\mu$m and {\it{ISO}} SWS and LWS in the remaining range. The {\it{ISO}} data
were smoothed to the resolution of {\it{Spitzer}} (R$\sim$600) and scaled to the {\it{Spitzer}}
fluxes by using maximum scaling factors of $\sim$20 $\%$. The resulting SEDs for both
objects are shown in Figure 2. 

The overall SEDs of IRAS 17423$-$1755 and IRAS 19343$+$2926 peak between 
40 and 60 $\mu$m. 
Notably, the absorption feature in the 2.5-3.5 $\mu$m region appears
strong in both objects, together with [CII] in emission at 158 $\mu$m; 
however, this latter feature might be Galactic residual background emission. 

The $\sim$2$-$180 $\mu$m SEDs in Figure 2 display a very cold bimodal  
continuum that can be interpreted in terms of thermal emission from the dust.
Obviously, it would correspond to several wavelength-dependent dust emissivities
with a certain range of dust temperatures, but we found that two main components
of the nebular dust, cold (strong) and hot (weaker), could roughly reproduce the general trend of
the observed overall energy distribution. We performed a multiple blackbody
fitting using the IRAF routine NLFIT on the F$_{\lambda}$ over $\lambda$
spectrum and obtained blackbody temperatures of 120 K and 965 K for IRAS
17423$-$1755 (in agreement with the values reported in \citet{Gauba04}) and
temperatures of 120 K and 965 K  in the case of IRAS
17423$-$1755. The fitted blackbody 
curves are shown in the F$_{\nu}$ over $\lambda$ spectra of 
Figure 2, where we also detail the fitting in the region of the 3.1 $\mu$m absorption feature. 
\citet{Bunzel09}
have recently reported difficulties in modeling the SEDs of heavily obscured O-rich
post-AGB stars by using a more sophisticated radiation transfer code for dusty
environments such as DUSTY. They needed to add amorphous carbon dust in their
DUSTY models in order to reproduce both the observed red continua and the
apparently weak 10 $\mu$m amorphous silicate absorption.\footnote{This problem
has also been found by \citet{Cerrigone09} and \citet{Gauba04}, who also needed to
introduce amorphous carbon grains to reproduce the observed SED of IRAS
17423$-$1755. This fact probably led these authors to identify the strong 3.1
$\mu$m absorption band as due to C$_{2}$H$_{2}$.} It is to be noted here that
the inclusion of amorphous carbon grains in the DUSTY models of O-rich post-AGB
stars does not necessarily mean that these sources are C-rich; their infrared
spectra show O-rich dust features only. Other more unusual O-rich dust
species with optical properties similar to those of the amorphous carbon grains could be
present in the circumstellar shells of heavily obscured O-rich post-AGB stars
(see Section 4), also giving a good fit to the observed SEDs of these stars (R.
Szczerba 2009, private communication). These difficulties prevent us from  proceeding further
 with this modeling, as the thermal continuum around 3 $\mu$m was found
to provide an adequate reference for the optical depth of the observed feature
and we are mainly interested in the analysis of the dust features present
in the new high-resolution  {\it{Spitzer}} spectrum. 

\section{Discussion}

\subsection{O-rich chemistry}
The most interesting and well-defined feature in the {\it{ISO}} spectrum of IRAS
17423$-$1755 is a clear absorption band at 3.1 $\mu$m, which we identify for the first
time as water ice (see below) present in the dust grains of the circumstellar
material. Similar detections are not very numerous,  the comparison object  being IRAS
19343$+$2926, a bipolar post-AGB star with an O-rich composition, another case that has
also been observed with {\it{ISO}}. The identification of this 3.1 $\mu$m absorption feature
as C$_{2}$H$_{2}$ by \citet{Gauba04} is erroneous, as the new
high-resolution  {\it{Spitzer}}/IRS spectrum confirms. Figure 3 displays the SED of IRAS 17423$-$1755 from 2 
to $\sim$16 $\mu$m together with the comparison O-rich post-AGB stars IRAS
19343$+$2926 and IRAS 17393$-$2727. No other feature characteristic of a
C-rich chemistry (e.g., SiC, PAHs, C$_{2}$H$_{2}$, HCN) in the circumstellar
envelope can be found in the observed  {\it{Spitzer}} spectrum. In particular, there
is no indication of the presence of C$_{2}$H$_{2}$ absorption at 13.7 $\mu$m and the key 
point here is that the C$_{2}$H$_{2}$ ro-vibrational lines at $\sim$3 and 14
$\mu$m are of similar strength ($\sim$20 $\%$ from the continuum; see e.g., Cernicharo
et al. (1999)). Thus, if the 3.1 $\mu$m absorption feature is due to C$_{2}$H$_{2}$,
then the C$_{2}$H$_{2}$ 13.7 $\mu$m feature should be detected in the
high-resolution  {\it{Spitzer}} spectrum presented here. The non-detection of this feature, or 
even the other small hydrocarbons detected in C-rich
post-AGB stars \citep{Cernicharo99,Cernicharo01} supports our identification of the 
3.1 $\mu$m absorption band with water ice.   

 {\it{Spitzer}}/IRS SL and SH spectra of IRAS 17423$-$1755 reveal for the first time a mid-IR feature that was not evident in the low S/N {\it{ISO}} spectrum \citep{Kraemer02, Gauba04} and that was not recognized by \citet{Cerrigone09} in their IRS/SL spectrum. A weak and broad 9.7 $\mu$m
absorption from amorphous silicates (O-rich) is present. This is shown in Figure 3, in comparison with the other well known O-rich post-AGB
stars IRAS 17393$-$2727 and IRAS 19343$+$2926. 

\citet{GarciaH07} discussed the presence of crystalline silicates in the  {\it{Spitzer}}/IRS spectrum of IRAS 17393$-$2727. They reported the presence of several weak crystalline silicate emission features in the 27-31 $\mu$m region, and crystalline silicate absorption features at shorter wavelengths. In particular, an absorption feature at 15.4 $\mu$m can be clearly observed in the IRS spectrum of this star presented in Figure 3.
Such crystalline silicate absorption/emission features cannot be claimed for definitely in either IRAS 17423$-$1755 or
IRAS 19343$+$2926 at the S/N level of our  {\it{Spitzer}}/IRS spectra. It is worth noticing here that these weak
crystalline silicate features have been observed in the circumstellar envelopes of O-rich evolved stars with a variety of strengths
and at slightly different wavelengths from source to source, depending
on the specific chemical composition and the size and density of the dust grains in the
circumstellar envelope \citep{GarciaH07}.

Our identification at 3.1 $\mu$m of water ice and the weak 9.7 $\mu$m amorphous silicates
absorption permit us to infer for the first time an O-rich chemistry for the
circumstellar envelope around IRAS 17423$-$1755. Figure 3 shows that a notable
similarity exists between the infrared spectra of the O-rich post-AGB star 
IRAS 19343$+$2926 and that of IRAS 17423$-$1755. In
addition, [NeII] nebular emission at 12.8 $\mu$m is also detected for the three 
objects shown in this figure, suggesting that the ionization of the circumstellar material may
have already started in these evolved stars. 

\subsection{Water ice in highly embedded evolved stars}

The conditions for condensation and properties of amorphous and crystalline H$_{2}$O
ice in astrophysical environments have been extensively reviewed by several authors
\citep{Leger79,Leger83,Baratta91,Kouchi94,Smith94}. Different forms of water ice can
be observed, depending mainly on temperature, pressure and the mechanisms of
deposition. For typical low pressure conditions present in the circumstellar
envelopes, temperatures higher than 150--170 K produce hexagonal ice, which remains
stable during further cooling. If deposition takes place  at lower temperatures
(between 110 and 130 K) cubic ice is formed, while at temperatures lower than 100--130 K
the resulting ice is amorphous \citep{Baratta91,Kouchi94}. 

In the circumstellar envelopes of evolved stars water ice can condense forming icy
mantles on dust grains which have previously condensed while the gas cooled down in
the expanding envelopes. In O-rich envelopes all the carbon atoms are thought to be locked
in CO molecules and the remaining O atoms are supposed to form H$_{2}$O aggregates.
Determining the occurrence and characterization of  water ice features is interesting
because ice in these mantles can provide important diagnostics of the physical
conditions in  circumstellar envelopes. Water ice has been observed in a
significant sample of AGB stars - in particular in OH/IR stars - some post-AGB stars
and a few PN \citep{Omont90,Eiroa89, Hoogzaad02,Molster01}. The specific conditions
for the formation of water ice in circumstellar envelopes around evolved stars have
been discussed and modeled by Dijkstra et al. (2003, 2006), who also
studied the trends that can be observed in the 3, 43 and 62 $\mu$m water ice features
during the stellar evolution from the AGB to the PN phase. 

According to \citet{Dijkstra06} it continues to be a puzzle why some stars form water
ice and not others. One clue these authors found might be the fact that the strength of 
the 3, 43 and 62 $\mu$m water ice features increases with the increasing initial mass of
the star. Their model calculations suggest that water ice features will be too weak to
be detectable for stars with zero age main sequence (ZAMS) masses lower than 5
M$_{\odot}$; the water ice features completely disappear for initial masses lower than
$\sim$ 3 M$_{\odot}$. A high mass loss rate also favors the detectability of the water
ice features, as well as large values of the mass loss rate to luminosity ratio. The
absence of a strong interstellar UV radiation field also preserves ices, as well as the
presence of a high density region that can provide shielding from an energetic radiation
field.

Both crystalline and amorphous water ice can form in circumstellar envelopes.  The
models developed by \citet{Dijkstra06} show that the shape of the spectral features is
very sensitive to the type of water ice aggregate formed. The 3.1 $\mu$m crystalline
water ice absorption feature is characterized by a sharp core with two unequally
strong shoulders. On the other hand, the amorphous water ice absorption feature is
broader, with no substructures, and displaced to shorter wavelengths. In the 30--100
$\mu$m region crystalline water ice shows two prominent and very broad emission
features near 43 and 62 $\mu$m. In the case of amorphous water ice, these latter
features are undetectable because they are broad and show almost no contrast with the
dust continuum.

Figure 4 details the 3.1 $\mu$m water ice band in IRAS 17423$-$1755 and in IRAS 19343$+$2926. In both cases
the band presents a sharp profile and shows the characteristic shoulders on both sides of the central core, indicating
that most of the water ice formed must be crystalline. This is demonstrated by the good match that can be appreciated among the observed profiles and the model displayed in the figure,
calculated for crystalline water ice deposited at a dust excitation temperature of 150 K by
\citet{Smith89}.

Additionally, the {\it{ISO}}
spectrum of IRAS 19343$+$2926 shows the presence of the water ice bands at 62 and 43
$\mu$m \citep{Dijkstra06}, confirming the crystalline nature of the water ice in this star. The presence of these
latter features cannot be excluded in the case of IRAS 17423$-$1755 because of the
poor quality of the {\it{ISO}} LWS spectrum (see Figure 2). A much higher S/N spectrum would be
needed in order to confirm/discard the detection of the crystalline water ice features
at the longer wavelengths.
The water ice features are providing us with valuable information for interpreting the
precise evolutionary stage of these nebulae. This point will be addressed in Section
4 together with the analysis of other observational properties such as the nebular
morphology and the presence of amorphous silicates.

\section{Evolutionary stage}

The available  {\it{HST}} optical images for IRAS 17423$-$1755 (Bobrowsky et al. 1995)
and IRAS 19343$+$2926 \citep{Trammell96} were retrieved from the  {\it{HST}}
archive in order to compare the morphological properties of both objects. These
optical images are presented in Figure 5, together with the  {\it{HST}} image of IRAS
17393$-$2727 (unpublished), showing that strong and highly collimated bipolar
outflows are observed in the three objects. Note that bipolar morphologies
are mainly found among type I PN, which are expected to be the descendants of
the evolution of the more massive AGB/post-AGB stars \citep{Corradi95}. 

The models by \citet{Dijkstra06} discussed in Section 3.2 have
shown that the conditions for the formation and prevalence of crystalline water
ices, like those observed in IRAS 17423$-$1755 and IRAS 19343$+$2926, would
imply high initial masses (at least higher than $\sim$3 M$_{\odot}$) for the
central star and the probable presence of high density structures such as a dusty
disk or torus, which could completely obscure the central star in the
optical/near-IR wavelength range. The existence of such dusty disk/torus
structures is confirmed in the optical images of both objects.
\citet{GarciaH07} have proposed that heavily obscured high-mass precursors of PN like
IRAS 17393$-$2727 may be already developing strong bipolar outflows, and it is
likely that a thick circumstellar disk/torus - where the crystallization of
water ice could take place - is surrounding the central post-AGB
star. The presence of a highly collimated bipolar outflow as well as a thick
circumstellar disk/torus that completely obscures the central source in IRAS
17393$-$2727 is now confirmed by the available  {\it{HST}} images shown in Figure
5.

\citet{Sylvester99} used the {\it{ISO}} spectra of a sample of OH/IR stars to analyze 
the IR characteristics of their circumstellar dust. They found that the optical depth
of the $\sim$10 and 18 $\mu$m amorphous silicate features increases with
increasing mass loss rate during the AGB.\footnote{Note that it is believed that
the mass-loss rate increases during the AGB phase \citep{vanderVeen88}}. 
The same results were found by \citet{Dijkstra06}, who modeled the
evolution of the 2--200 $\mu$m SED for a 5 M$_{\odot}$ (ZAMS) star evolving from
the early AGB until the early PN stage. At the end of the AGB, the circumstellar
envelope detaches from the star, becoming optically thin and causing the
amorphous silicate features to disappear. This infrared evolutionary sequence
from the AGB to the PN stage has been observationally confirmed by \citet{GarciaH07}
using  {\it{Spitzer}} spectra of massive O-rich AGB/post-AGB stars. 
Regarding the evolutionary state of IRAS 17423$-$1755, it is likely that this star 
is more evolved than IRAS 19343$+$2926 because the
amorphous silicate absorption and the water ice band are weaker
\citep{Dijkstra06}. IRAS 17393$-$2727 could be the least evolved object of the
three because it has no optical/near-IR counterpart (the central star is
not detected even in the K-band). However, in the case of IRAS 19343$+$2926 and
IRAS 17423$-$1755, the central star has already reappeared in the
optical/near-IR range. The presence of [NeII] nebular emission could provide
evidence on the onset of the circumstellar envelope ionization, although shock
excitation in the high velocity outflows cannot be excluded. 

The SED of IRAS 17423$-$1755 is similar to those of extreme OH/IR AGB stars,
which exhibit absorption features from amorphous silicates together with
crystalline silicate features that alternate between emission and absorption
depending on the specific physical and chemical properties of the circumstellar
dust grains \citep{Sylvester99}. We think that the presence of crystalline silicates in the 13-35 $\mu$m region in IRAS 17423$-$1755 and IRAS 19343$+$2926 cannot be excluded but that higher S/N spectra would be needed in order to confirm or discard such a presence.

In summary, the bipolar morphology, the detection of OH maser emission and
crystalline water ice as well as the 9.7 $\mu$m absorption from amorphous silicates present in the  {\it{Spitzer}}
spectrum indicate that IRAS 17423$-$1755 is a massive O-rich post-AGB star
similar to IRAS 19343$+$2926 and IRAS 17393$-$2727. The three sources represent
a link between massive OH/IR AGB stars and bipolar type I PN. 

\section{Conclusions}

An O-rich chemistry for the circumstellar envelope around the post-AGB object
IRAS 17423$-$1755 is here confirmed, despite a previous classification as
C-rich. This result is based on the detection of a weak and broad 9.7 $\mu$m
amorphous silicate absorption in the high-resolution  {\it{Spitzer}}/IRS spectrum. The complete lack
of C-rich mid-IR features (in particular C$_{2}$H$_{2}$ at 13.7 $\mu$m) supports our
identification of the strong 3.1 $\mu$m absorption band seen in the {\it{ISO}} spectrum
as due to water ice as well as our O-rich classification for IRAS 17423$-$1755.

IRAS 17423$-$1755, IRAS 19343$+$2926 and IRAS 17393$-$2727 present clear
evidences of the presence of a circumstellar disk or torus, where the conditions
would be very similar to those found in less evolved and more embedded OH/IR
stars. A recent strong mass-loss event has been reported in the case of IRAS
19343$+$2926 \citep{Alcolea07}, which would favor this scenario. Water ice and crystalline silicates would preferentially form in the outer region of the inner
torus, where low temperature conditions and shielding from the central star
would allow  a favorable rate of crystallization to take place. Both in IRAS
17423$-$1755 and IRAS 19343$+$2926, the ice band at 3.1 $\mu$m is sharp and
presents substructures, and  comparison with models of water ice growth
around evolved stars \citep{Smith89, Dijkstra06} allows us to confirm that the
ice is mostly in a crystalline state.

The morphological properties, detection of OH maser emission, and the
 {\it{Spitzer}}/IRS spectra observed in IRAS 17423$-$1755 are similar to those of
the O-rich post-AGB stars IRAS 19343$+$2926 and IRAS 17393$-$2727, allowing us
to interpret the evolutionary stage of IRAS 17423$-$1755 as belonging to an
intermediate stage between those OH/IR stars with extreme outflows and highly
bipolar type I PN.

\begin{acknowledgements}

M. M. and A. U. acknowledge financial support from the Spanish  Ministry 
of Science and Innovation (MICINN) through grant
AYA 2009-14648-02 and from the Xunta de Galicia through grants INCITE09
E1R312096ES and INCITE09 312191PR, all of them partially supported by E.U. FEDER
funds. D. A. G. H. and A. M. acknowledge support for this work provided by the
MICINN under the 2008 Juan de La Cierva Program and under grant AYA-2007-64748.
 This work is based on spectra obtained with {\it{ISO}}, an ESA project with instruments funded by ESA Member States 
(especially the PI countries: France, Germany, The Netherlands and the UK) 
and with participation of ISAS and NASA;
and  {\it{Spitzer}}, a NASA's Great Observatories Program.

\end{acknowledgements}

\clearpage

\begin{figure}
  \centering
\includegraphics[width=0.95\textwidth]{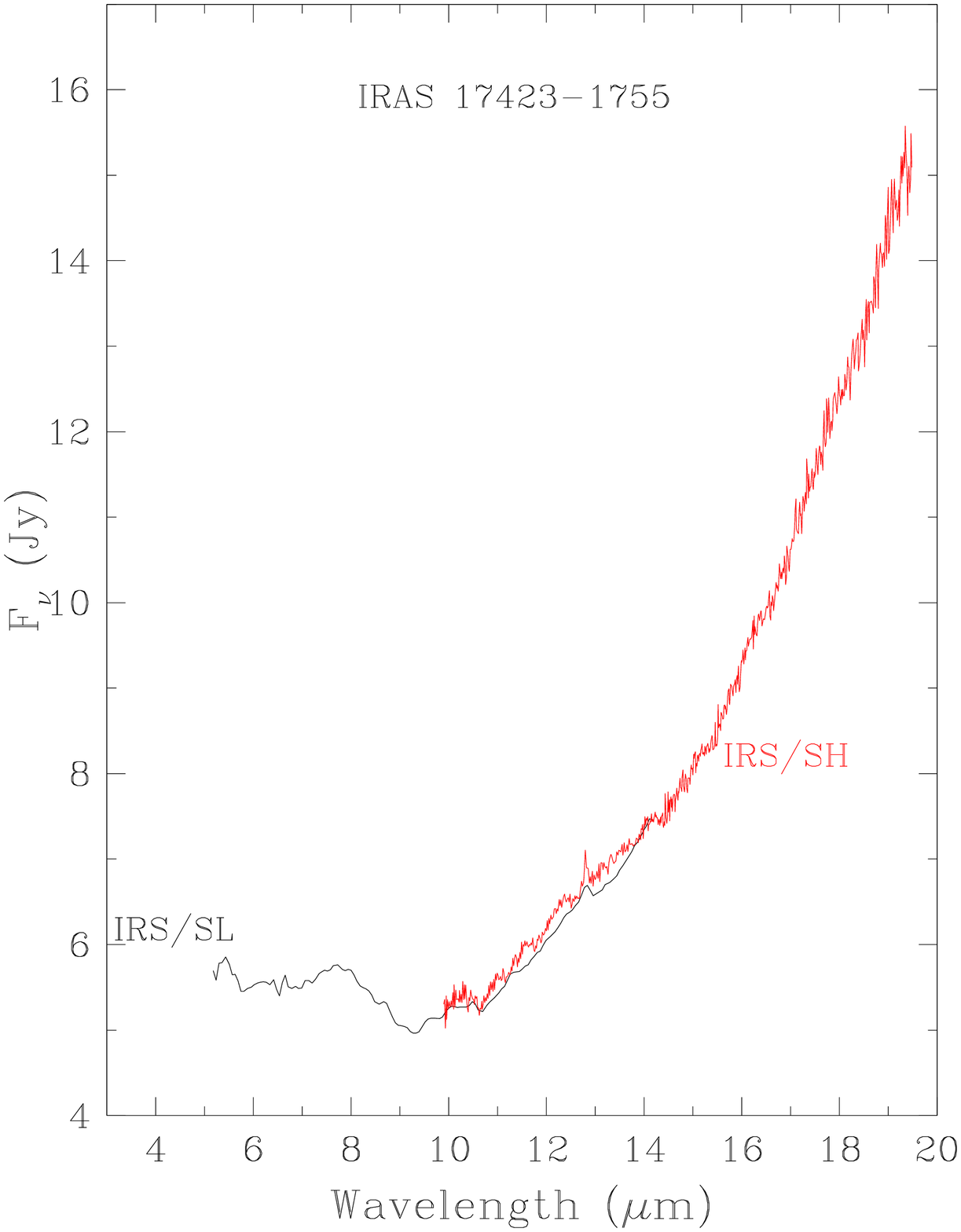}
   \caption{Overlap region between the  {\it{Spitzer}}/IRS SL spectrum and SH
of IRAS 17423$-$1755 showing the agreement between both spectra ($\leq$5\%).}
\label{fig1}
   \end{figure}
\clearpage

\begin{figure}
  \centering
\includegraphics[width=0.95\textwidth]{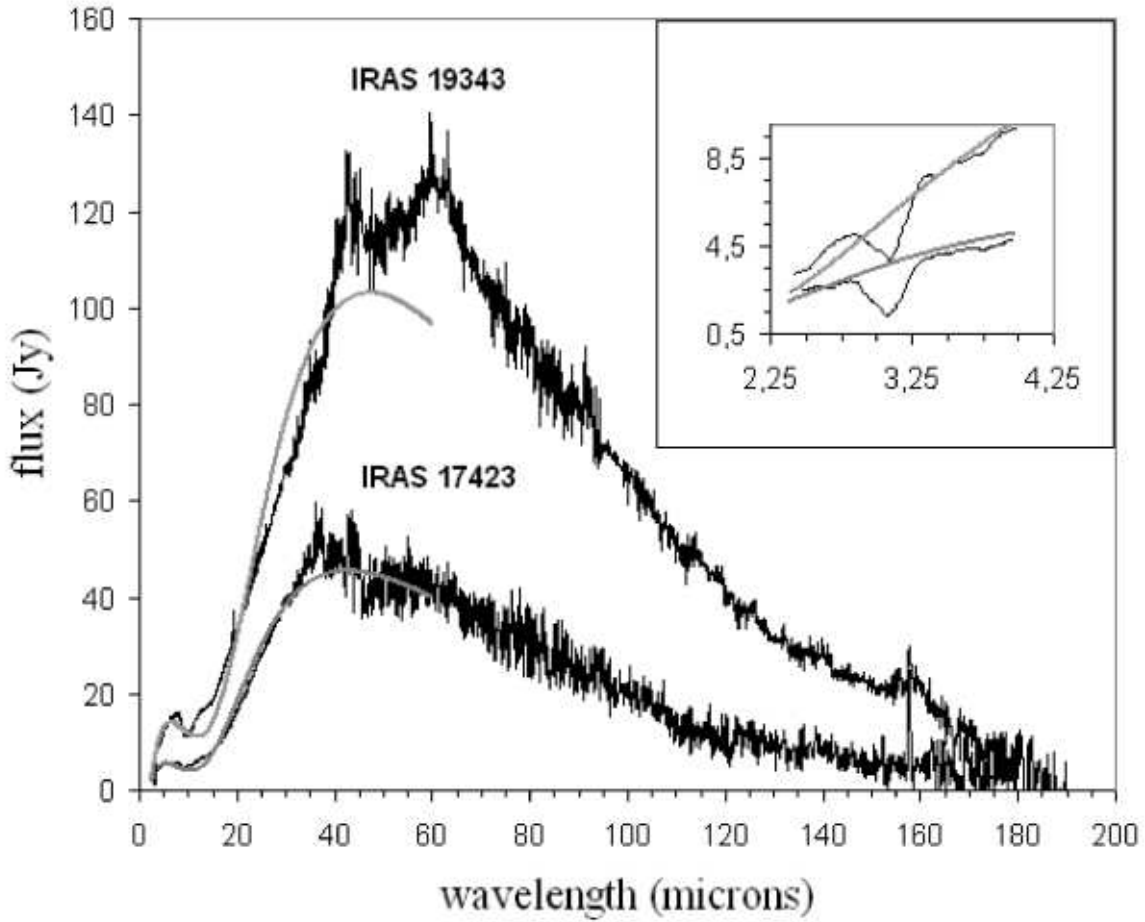}
   \caption{SED from $\sim$2 to 180 $\mu$m for IRAS 17423$-$1755
and IRAS 19343$+$2926. The SEDs were constructed by combining  {\it{Spitzer}} and {\it{ISO}} data (see the text for details). The blackbody curves fitted to the spectra of both stars are also shown, corresponding to dust temperatures of 120 K and 965 K for IRAS 17423$-$1755, and 107 K and 800 K for IRAS 19343$+$2926. The spectral region around 3 $\mu$m is detailed in the small window.}
      \label{fig2}
   \end{figure}
\clearpage

\begin{figure}
  \centering
\includegraphics[angle=-90,width=0.95\textwidth]{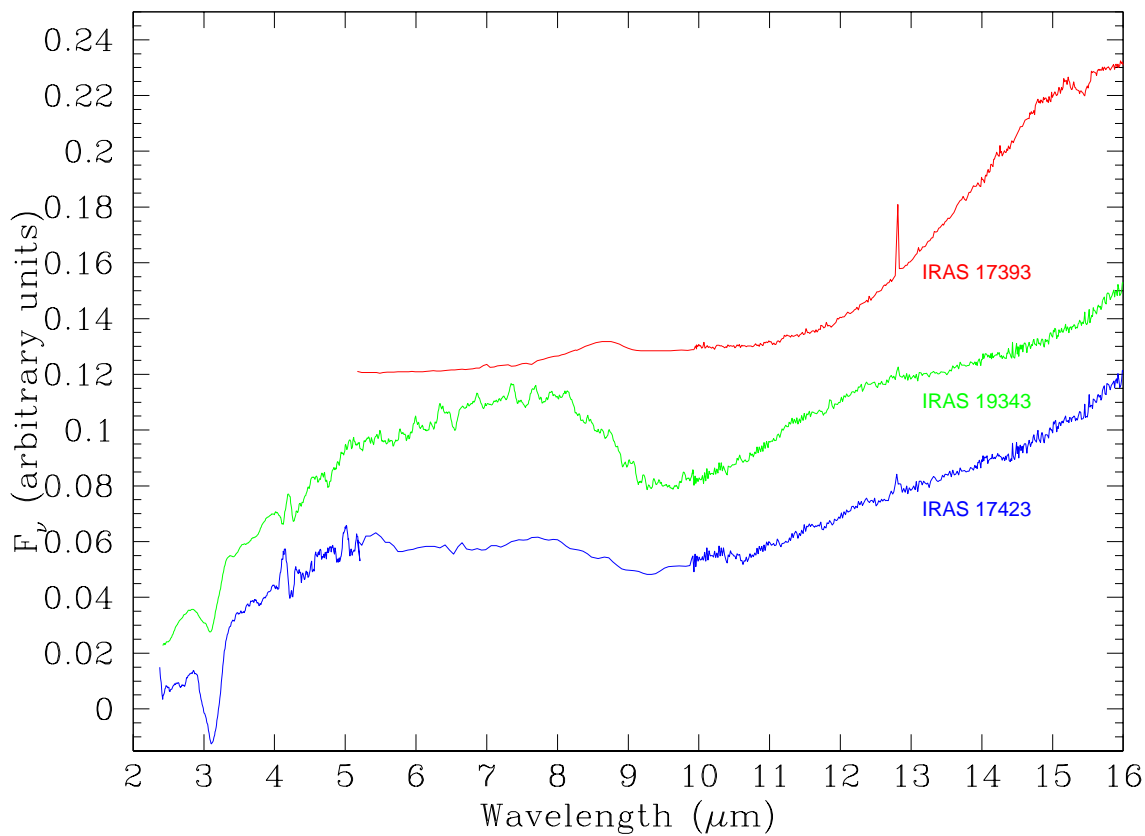}
   \caption{ {\it{Spitzer}}/IRS and {\it{ISO}} SWS spectra from $\sim$2 to 16 $\mu$m for IRAS
17423$-$1755 in comparison with those of the O-rich post-AGB stars IRAS 19343$+$2926 and
IRAS 17393$-$2727. This figure illustrates the presence of the broad 9.7 $\mu$m
absorption from amorphous silicates (O-rich) and the absence of C$_{2}$H$_{2}$ absorption at 13.7 $\mu$m.
[NeII] 12.8 $\mu$m nebular emission is clearly detected in
the three sources.}
  \label{fig3}
   \end{figure}
\clearpage

%

%
\begin{figure}
  \centering
 \includegraphics[width=0.80\textwidth]{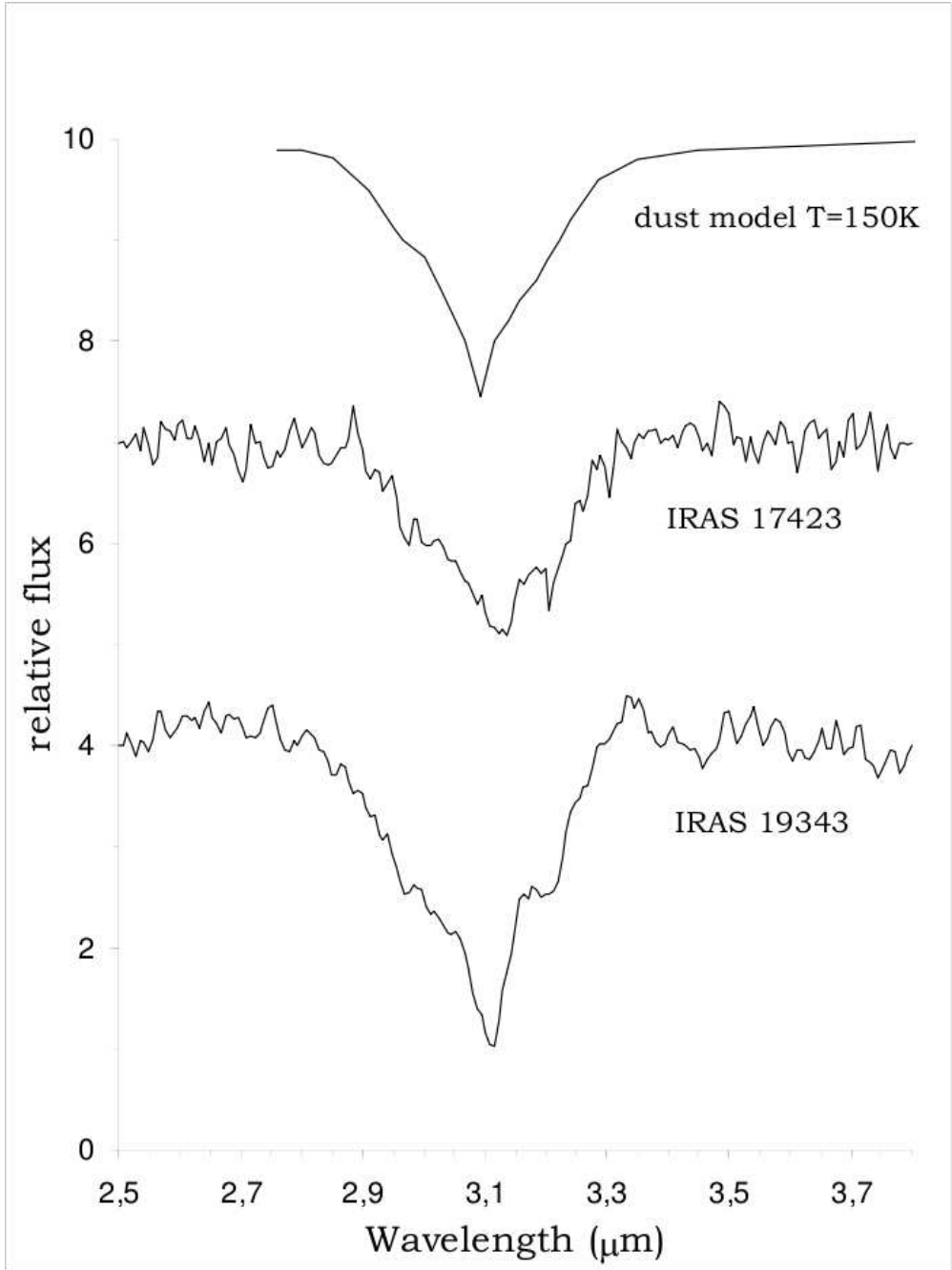}
   \caption{Observed profile of the 3.1 microns water ice feature (continuum divided) in IRAS 17423$-$1755 and IRAS 19343$+$2926 and the theoretical profile derived by Smith et al. 1989 for crystalline water ice-coated silicate grains at 150 K.
             }
        \label{fig4}
   \end{figure}

\clearpage

\begin{figure} 
\begin{center}$
\begin{array}{ccc}
\includegraphics[angle=0,scale=4.7]{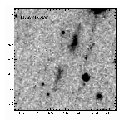} &
\includegraphics[angle=0,scale=4.7]{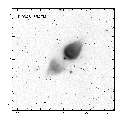}
\includegraphics[angle=0,scale=4.7]{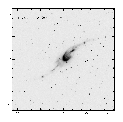}
\end{array}$
\end{center}
\caption{Optical  {\it{HST}} images retrieved from the  {\it{HST}} archive for the O-rich
post-AGB stars IRAS 17393$-$2727 (left panel, unpublished), IRAS 19343$+$2926
(middle panel, original images reported by Trammell \& Goodrich(1996)) and IRAS
17423$-$1755 (right panel; original images reported by Bobrowsky et al. 1995).
North is up and east to the left; axes are in arcsecs. The optical filters (F606W, F547M, F555W) are
indicated in each panel. The images are shown on a logarithmic scale. Note the
strong and highly collimated bipolar outflows seen in the three objects.
}
\label{fig5} 
\end{figure}


\begin{thebibliography}{}
\bibitem[Alcolea et al.(2007)]{Alcolea07} Alcolea, J., Neri, R., \& Bujarrabal, V. 2007, A\&A, 468, L41-L44
\bibitem[Baratta et al.(1991)]{Baratta91} Baratta, G. A., Spinella, F., Leto, G., Strazzulla, G., \& Foti, G. 1991, A\&A, 252, 421
\bibitem[Bobrowsky et al.(1995)]{Bobrowski95} Bobrowsky, M., Zijlstra, A. A., Grebel, E. K., Tinney, C. G., te Lintel Hekkert, P., van de Steene, G. C., Likkel, L., \& Bedding, T. R. 1995, ApJ, 446, L89
\bibitem[Bunzel et al.(2009)]{Bunzel09} Bunzel, F., Garc\'\i a -Hern\'andez, D. A., Engels, D., Perea-Calder\'on, J. V., \& Garc\'\i a-Lario, P. 2009, in ASP Conf. Ser. 418, AKARI, A Light to Illuminate the Misty Universe, ed. T. Osaka et al. (San Francisco, CA: ASP), 431
\bibitem[Cernicharo et al.(1999)]{Cernicharo99} Cernicharo, J., Yamamura, I., Gonz\'alez-Alfonso, E., de Jong, T., Heras, A., Escribano, R., \& Ortigoso, J. 1999, ApJ, 526, L41
\bibitem[Cernicharo et al.(2001)]{Cernicharo01} Cernicharo, J., Heras, A. M., Tielens, A. G. G. M., Pardo, J. R., Herpin, F., Gu\'elin, M., \& Waters, L. B. F. M. ApJ, 546, L123

\bibitem[Cerrigone et al.(2009)]{Cerrigone09} Cerrigone, L., Hora, J. L., Umana, G., \& Trigilio, C. 2009, ApJ, 703, 385
\bibitem[Corradi \& Schwarz(1995)]{Corradi95} Corradi, R. L. M., \& Schwarz, H. E. 1995, A\&A, 293, 871
\bibitem[de Graauw et al.(1996)]{deGraauw96} de Graauw, T.,  Haser, L. N.,  Beintema, D. A.,  Roelfsema, P. R.,  van Agthoven, H., Barl, L.,  Bauer, O. H.,  Bekenkamp, H. E. G. et al . 1996, A\&A, 315, L49
\bibitem[Dijkstra et al.(2003)]{Dijkstra03} Dijkstra, C., Dominik, C., Hoogzaad, S. N., de Koter, A., \&  Min, M. 2003, A\&A, 401, 599
\bibitem[Dijkstra et al.(2006)]{Dijkstra06} Dijkstra, C., Dominic, C., Bouwman, J., de Koter, A. 2006, A\&A, 449, 1101
\bibitem[Eiroa et al.(1983)]{Eiroa83} Eiroa, C., Hefele, H., \& Qian, Z.-Y. 1983, A\&AS, 54, 309
\bibitem[Eiroa \& Hodapp(1989)]{Eiroa89} Eiroa, C. \& Hodapp, K.-W. 1989, A\&A, 223, 271
\bibitem[Gauba \& Parthasarathy(2004)]{Gauba04} Gauba, G.\& Parthasarathy, M. 2004, A\&A, 417, 201
\bibitem[Garc\'\i a-Hern\'andez et al.(2007)]{GarciaH07} Garc\'\i a-Hern\'andez, D. A., Perea-Calder\'on, J.V., Bobrowsky, M., \&  Garc\'\i a-Lario, P. 2007, ApJ, 666, L33
\bibitem[Garc\'\i a-Hern\'andez et al.(2009)]{GarciaH09} Garc\'\i a-Hern\'andez, D. A., Bunzel, F., Engels, D., Perea-Calder\'on, J. V., \& Garc\'\i a-Lario, P. 2009, in ASP Conf. Ser. 418, AKARI, A Light to Illuminate the Misty Universe, ed. T. Osaka et al. (San Francisco, CA: ASP), 435
\bibitem[Garc\'\i a-Lario et al.(1999)]{GarciaLario99} Garc\'\i a-Lario, P., Manchado, A., Ulla, A., Manteiga, M. 1999, ApJ, 513, 941
\bibitem[Herwig(2005)]{Herwig05} Herwig, F. 2005, ARA\&A, 43, 435
\bibitem[Higdon et al.(2004)]{Higdon04} Higdon, S. J. U., Devost, D., Higdon, J. L., Brandl, B. R., Houck, J. R., Hall, P., Barry, D., Charmandaris, V. et al. 2004, PASP, 116, 975 
\bibitem[Hoogzaad et al.(2002)]{Hoogzaad02} Hoogzaad, S. N., Molster, F. J., Dominik, C., Waters, L. B. F. M., Barlow, M. J., \& de Koter, A. 2002, A\&A, 547
\bibitem[Ivezi\'c \& Elitzur(1997)]{Ivezic97} Ivezi\'c, Z., \& Elitzur, M. 1997, MNRAS, 287, 799
\bibitem[Kouchi et al.(1994)]{Kouchi94} Kouchi, A., Yamamoto, T., Kozasa, T., Kuroda, T., \& Greenberg, J. M. 1994, A\&A, 290, 1009
\bibitem[Kraemer et al.(2002)]{Kraemer02} Kraemer, K. E., Sloan, G. C., Price, S. D., Walker, H. J. 2002, ApJS, 140, 389
\bibitem[Leger et al.(1979)]{Leger79} Leger, A., Klein, J., de Cheveigne, S., Guinet, C., Defourneau, D., \& Belin, M. 1979, A\&A,  79, 256
\bibitem[Leger et al.(1983)]{Leger83} Leger, A., Gauthier, S., Defourneau, D., \& Rouan, D. 1983, A\&A, 117, 164
\bibitem[Molster et al.(2001)]{Molster01} Molster, F. J., Lim, T. L., Sylvester, R. J., Waters, L. B. F. M., Barlow, M. J., Beintema, D. A., Cohen, M., Cox, P. et al., 2001, A\&A, 372, 165
\bibitem[Omont et al.(1990)]{Omont90} Omont, A., Forveille, T., Moseley, S. H., Glaccum, W. J., Harvey, P. M., Likkel, L., Loewenstein, R. F., \& Lisse, C. M. 1990, ApJ, 355, L27
\bibitem[Parthasarathy \& Pottasch(1989)]{ParthaPottasch89} Parthasarathy, M., \& Pottasch, S. R. 1989, A\&A, 225, 521
\bibitem[Riera et al.(1995)]{Riera95} Riera, A., Garc\'\i a-Lario, P., Manchado, A., Pottasch, S. R., \& Raga, A. C. 1995, A\&A, 302, 137
\bibitem[S\'anchez-Contreras \& Sahai (2001)]{SanchezContreras01} S\'anchez-Contreras, C. \& Sahai, R. 2001, ApJ 553,L173
\bibitem[Smith et al.(1989)]{Smith89} Smith, R. G., Sellgren, K., \& Tokunaga, A. T. 1989, ApJ 344, 413
\bibitem[Smith et al.(1994)]{Smith94} Smith, R. G., Robinson, G., Hyland, A. R. , \& Carpenter, G. L. 1994,  MNRAS, 271, 481
\bibitem[Sylvester et al.(1999)]{Sylvester99} Sylvester, R. J., Kemper, F., Barlow, M. J., de Jong, T., Waters, L. B. F. M., Tielens, A. G. G. M., \& Omont, A. 1999, A\&A, 352, 587
\bibitem[Trammell \& Goodrich(1996)]{Trammell96} Trammell, S. R., \& Goodrich, R. W. 1996, ApJ, 468, L107
\bibitem[van der Veen \& Habing(1988)]{vanderVeen88} van der Veen, W. E. C. J., \& Habing, H. J. 1988, A\&A, 195, 125
\bibitem[van Winckel(2003)]{vanWinckel03} van Winckel, H. 2003, ARA\&A, 41, 391
\end{thebibliography}
\end{document}